\documentclass[12pt]{article}
\usepackage[totalwidth=460truept,totalheight=628truept]{geometry}
\usepackage{latexsym,graphicx,amsfonts,amssymb,amsmath}
\usepackage[hypertexnames=false,hidelinks]{hyperref}

\def\theequation{\arabic{section}.\arabic{equation}}

\renewcommand{\theequation}{\thesection.\arabic{equation}}
\global\arraycolsep=1truept
\renewcommand{\theequation}{\arabic{section}.\arabic{equation}}

\begin{document}

\null

\vskip1truecm

\begin{center}
{\huge \textbf{High-Order Corrections}}

\vskip.8truecm

{\huge \textbf{to Inflationary Perturbation Spectra}}

\vskip.8truecm

{\huge \textbf{in Quantum Gravity}}

\vskip1truecm

\textsl{Damiano Anselmi}

\vskip .1truecm

\textit{Dipartimento di Fisica \textquotedblleft Enrico Fermi", Universit%
\`{a} di Pisa}

\textit{Largo B. Pontecorvo 3, 56127 Pisa, Italy}

\textit{and INFN, Sezione di Pisa,}

\textit{Largo B. Pontecorvo 3, 56127 Pisa, Italy}

damiano.anselmi@unipi.it

\vskip1truecm

\textbf{Abstract}
\end{center}

We compute the inflationary perturbation spectra and the quantity $r+8n_{T}$
to the next-to-next-to-leading log order in quantum gravity with purely
virtual particles (which means the theory $R+R^{2}+C^{2}$ with the fakeon
prescription/projection for $C^{2}$). The spectra are functions of the
inflationary running coupling $\alpha (1/k)$ and satisfy the cosmic
renormalization-group flow equations, which determine the tilts and the
running coefficients. The tensor fluctuations receive contributions from the
spin-2 fakeon $\chi _{\mu \nu }$ at every order of the expansion in powers
of $\alpha \sim 1/115$. The dependence of the scalar spectrum on the $\chi
_{\mu \nu }$ mass $m_{\chi }$, on the other hand, starts from the $\alpha
^{2}$ corrections, which are handled perturbatively in the ratio $m_{\phi
}/m_{\chi }$, where $m_{\phi }$ is the inflaton mass. The predictions have
theoretical errors ranging from $\alpha ^{4}\sim 10^{-8}$ to $\alpha
^{3}\sim 10^{-6}$. Nontrivial issues concerning the fakeon projection at
higher orders are addressed.

\vfill\eject

\section{Introduction}

\label{intro}\setcounter{equation}{0}

Quantum field theory, through the constraints of locality, renormalizability
and unitarity, is equipped with powerful tools to overcome the arbitrariness
of classical theories. Following a line of reasoning similar to the one that
lead to the standard model of particle physics, a basically unique quantum
field theory of gravity emerges from the constraints just mentioned \cite%
{LWgrav}. It is rooted into the concept of purely virtual particle, or
fakeon, which can be introduced by adopting a new quantization prescription
for the poles of a free propagator, alternative to the Feynman $i\epsilon $
one. The degrees of freedom associated with those poles can then be
projected away from the physical spectrum consistently with unitarity.

When the matter sector is switched off, the quantum field theory of gravity
propagates a triplet made of the graviton, a massive scalar field $\phi $
(the inflaton) and a massive spin-2 fakeon $\chi _{\mu \nu }$. The best
chances to test its predictions, possibly within years, are offered by
primordial cosmology. Applying the idea of purely virtual particle to
inflation, the mass $m_{\chi }$ of $\chi _{\mu \nu }$ is bound to be larger
than $m_{\phi }/4$, where $m_{\phi }$ is the $\phi $ mass, and the
tensor-to-scalar ratio $r$ is predicted within less than one order of
magnitude \cite{ABP}. The measurement of $r$, which is expected to be
achieved in the next future, will fix $m_{\chi }$ and determine the theory
completely. With some technical developments, primordial cosmology might
turn into an arena for precision tests of quantum gravity.

Calculations involving purely virtual particles do not require much more
effort than usual (see \cite{UVQG,Absograv} for renormalization constants,
widths and absorptive parts, \cite{ABP,CMBrunning} for inflationary
cosmology). Actually, various similarities between primordial cosmology and
high-energy physics allow us to import techniques from quantum field theory
to boost the computations of higher-order corrections. In this context, it
is convenient to formulate inflation as a \textquotedblleft
cosmic\textquotedblright\ renormalization-group (RG) flow, with a
\textquotedblleft fine structure constant\textquotedblright\ $\alpha \sim
1/115$ \cite{CMBrunning}. As proved in \cite{CMBrunning}, the power spectra
satisfy RG flow equations in the superhorizon limit. The computations of
tilts, running coefficients and higher-order corrections can be simplified
by resumming the leading and subleading logs and expressing the spectra $%
\mathcal{P}(k)$ as functions of the running coupling $\alpha (1/k)$.

In this paper, we extend the results of \cite{CMBrunning} by one order of
magnitude. Specifically, we compute the RG\ improved power spectra of the
curvature perturbation $\mathcal{R}$ and the tensor fluctuations to the
next-to-next-to-leading log (NNLL)\ order in the superhorizon limit. Tilts
and running coefficients follow straightforwardly from the flow equations.
We also compute the quantity $r+8n_{T}$, where $n_{T}$ denotes the tensor
tilt. A coincidence makes the first contribution to $r+8n_{T}$ vanish in
quantum gravity with fakeons \cite{ABP}, as well as in several other models
explored in the literature \cite{scenarios}. However, no argument suggests
that it should vanish identically. Our results show that it is a factor $%
\alpha $ smaller than naively expected, but definitely not zero.

The outcomes highlight the effects of the fakeon $\chi _{\mu \nu }$ on the
spectra. While the spectrum of the tensor fluctuations depends on $m_{\chi }$
at every order of the expansion in powers of\ $\alpha $, the spectrum of the
scalar fluctuations is affected by $\chi _{\mu \nu }$ only from the NNLL
order onwards. The calculation of the NNLL correction to the scalar spectrum
involves nontrivial aspects of the fakeon projection, which lead in general
to a nonlocal Mukhanov-Sasaki action. A way to avoid these difficulties is
to expand in powers of the ratio $m_{\phi }/m_{\chi }$. So doing, the result
contains a function of $\xi =m_{\phi }^{2}/m_{\chi }^{2}$, which we
calculate to order $\xi ^{9}$. The first nine contributions suggest that the
expansion is asymptotic. Nevertheless, it gives precise predictions for $%
m_{\chi }^{2}>2m_{\phi }^{2}$ and fair predictions for $2m_{\phi
}^{2}>m_{\chi }^{2}>m_{\phi }^{2}$. It is not very helpful in the rest of
the $\xi $ range, which is $m_{\phi }>m_{\chi }>m_{\phi }/4$.

In the case of the tensor spectrum, instead, no expansion in $\xi $ is
necessary, since $\xi $ can be handled exactly to the NNLL order included.
All in all, the relative theoretical errors of the physical predictions we
obtain in this paper range from $\alpha ^{4}\sim 10^{-8}$ to $\alpha
^{3}\sim 10^{-6}$. We also perform a number of independent checks of the RG\
equations.

Earlier calculations of the running of spectral indices in various scenarios
(in models without fakeons) can be found in \cite{run1} and calculations of
subleading corrections can be found in \cite{run2}. The cosmic RG flow
offers a way to upgrade the techniques used previously and provides a better
insight into the structures of the spectra of primordial fluctuations.

The paper is organized as follows. In section \ref{betaf} we briefly review
quantum gravity with fakeons and the cosmic RG flow, its beta function and
the\ evolution equations satisfied by the RG improved spectra. In section %
\ref{R2} we study the scalar and tensor spectra to the NNLL\ order in the
limit $m_{\chi }\rightarrow \infty $, where the theory tends to the
Starobinsky $R+R^{2}$ model \cite{starobinsky,vilenkin}. In section \ref%
{tensorQG} we calculate the spectrum of the tensor fluctuations to the NNLL\
order in quantum gravity, while in section \ref{scalarocm} we do the same
for the spectrum of the curvature perturbation $\mathcal{R}$. In section \ref%
{predictions} we collect the physical predictions and estimate their
theoretical errors. Section \ref{conclusions} contains the conclusions,
while appendix \ref{formulas} collects some reference formulas that are
useful for the calculations.

\section{Inflationary beta function and cosmic RG\ flow}

\label{betaf}\setcounter{equation}{0}

Quantum gravity with fakeons \cite{LWgrav} is described by a triplet made of
the graviton, a massive scalar $\phi $ (the inflaton) and a massive spin-2
fakeon $\chi _{\mu \nu }$. It can be formulated starting from the classical
action%
\begin{equation}
S_{\text{QG}}=-\frac{1}{16\pi G}\int \mathrm{d}^{4}x\sqrt{-g}\left( R+\frac{1%
}{2m_{\chi }^{2}}C_{\mu \nu \rho \sigma }C^{\mu \nu \rho \sigma }\right) +%
\frac{1}{2}\int \mathrm{d}^{4}x\sqrt{-g}\left( D_{\mu }\phi D^{\mu }\phi
-2V(\phi )\right) ,  \label{sqgeq}
\end{equation}%
where 
\begin{equation}
V(\phi )=\frac{3m_{\phi }^{2}}{32\pi G}\left( 1-\mathrm{e}^{\phi \sqrt{16\pi
G/3}}\right) ^{2}  \label{staropote}
\end{equation}%
is the Starobinsky potential and $m_{\phi }$ and $m_{\chi }$ are the masses
of $\phi $ and $\chi _{\mu \nu }$, respectively. For convenience, the
cosmological term and the matter sector are switched off. The action (\ref%
{sqgeq}) does not contain the fakeon $\chi _{\mu \nu }$ explicitly. We
identify it from the appropriate pole of the two-point function of the
metric fluctuation around flat space.

The theory is renormalizable, because (\ref{sqgeq}) is equivalent to the
higher-derivative action%
\begin{equation}
S_{\text{geom}}(g,\Phi )=-\frac{1}{16\pi G}\int \mathrm{d}^{4}x\sqrt{-g}%
\left( R+\frac{1}{2m_{\chi }^{2}}C_{\mu \nu \rho \sigma }C^{\mu \nu \rho
\sigma }-\frac{R^{2}}{6m_{\phi }^{2}}\right)  \label{Sgeom}
\end{equation}%
up to a standard, nonderivative field redefinition. Once the cosmological
term is reinstated (\ref{Sgeom}) is manifestly renormalizable by power
counting \cite{stelle}.

If all the degrees of freedom, defined by expanding the metric around flat
space, are quantized by means of the Feynman $i\epsilon $ prescription,
Stelle's theory \cite{stelle} is obtained, which violates unitarity. The
reason is that the $\chi _{\mu \nu }$ propagator is multiplied by the wrong
sign, so the Feynman prescription generates a ghost. The problem can be
avoided by adopting the Feynman prescription just for the graviton and the
inflaton, while quantizing $\chi _{\mu \nu }$ by means of the fakeon
prescription \cite{LWgrav}. So doing, $\chi _{\mu \nu }$ becomes a spin-2
purely virtual particle \cite{wheelerons} and can be projected away from the
physical spectrum consistently with unitarity. The informations we need on
the fakeon projection are given in sections \ref{tensorQG} and \ref%
{scalarocm} (for a more detailed review, see \cite{classicization}).
Important things to know are that, once we make the choices just stated:

1) the theory is unitary (see \cite{LWFormulation,LWUnitarity} for the
analysis of bubble and triangle diagrams in related models, \cite{fakeons}
for the proof to all orders, \cite{UVQG,Absograv} for the absorptive parts
in quantum gravity);

2)\ the theory remains renormalizable by power counting \cite{LWgrav,fakeons}%
; its beta functions and renormalization constants coincide with those of
the Euclidean version \cite{betaHD};

3) the true classical limit is not described by either (\ref{sqgeq}) or (\ref%
{Sgeom}), which are unprojected; it is obtained by \textquotedblleft
classicizing\textquotedblright\ quantum gravity as explained in \cite%
{classicization,FLRW};

4) the classicization is more challenging when the metric is expanded around
nontrivial backgrounds rather than flat space; yet, a lucky coincidence,
crucial for cosmology, makes the degrees of freedom decouple from one
another at the quadratic level when the background is the FLRW metric in the
de Sitter limit \cite{ABP}; this fact allows us to hook the fakeon
projection on a curved background to the flat-space one and proceed
perturbatively from there.

It can be shown \cite{ABP} that the procedure described in point 4) works
under the consistency condition $m_{\chi }>m_{\phi }/4$, which puts a lower
bound on the mass of the fakeon $\chi _{\mu \nu }$ with respect to the mass
of the inflaton $\phi $. Note that a bound of this type is nonperturbative,
if viewed from the expansion around flat space. Combining the constraints
coming from high-energy physics (that is to say, the requirements of
locality, renormalizability and unitarity, which make (\ref{Sgeom})
essentially unique \cite{LWgrav}) with those coming from cosmology, a very
predictive theory emerges, to the extent that the tensor-to-scalar ratio $r$
is determined within less than an order of magnitude, even before knowing
the value of $m_{\chi }$ \cite{ABP}.

Another important property is that the quantum field theory of gravity does
not predict other degrees of freedom besides the curvature perturbation $%
\mathcal{R}$ and the tensor fluctuations, when the matter sector is switched
off. Indeed, the fakeon projection eliminates the possibility of having
additional scalar and tensor perturbations, as well as the vector
perturbations.

In the rest of this section we recall the main features of the cosmic RG
flow introduced in \cite{CMBrunning} and the RG equations satisfied by the
spectra. Given the Friedmann equations and the $\phi $ equation%
\begin{equation}
\dot{H}=-4\pi G\dot{\phi}^{2},\qquad H^{2}=\frac{4\pi G}{3}\left( \dot{\phi}%
^{2}+2V(\phi )\right) ,\qquad \ddot{\phi}+3H\dot{\phi}=-V^{\prime }(\phi ),
\label{frie}
\end{equation}%
where $H=\dot{a}/a$ is the Hubble parameter, we define the coupling\footnote{%
For the purposes of this paper, we can assume $\dot{\phi}>0$. When $\dot{\phi%
}<0$, $\alpha $ becomes negative and the last equality of (\ref{alf}) must
be replaced by $\alpha =-$ $\sqrt{-\dot{H}}/\sqrt{3H^{2}}$.}%
\begin{equation}
\alpha =\sqrt{\frac{4\pi G}{3}}\frac{\dot{\phi}}{H}=\sqrt{-\frac{\dot{H}}{%
3H^{2}}}.  \label{alf}
\end{equation}%
Eliminating $V$ and $\dot{\phi}$ by means of the first two equations of (\ref%
{frie}) and $\ddot{\phi}$ from the last equation, it is easy to show that $%
\alpha $ satisfies%
\begin{equation}
\dot{\alpha}=m_{\phi }\sqrt{1-\alpha ^{2}}-H(2+3\alpha )\left( 1-\alpha
^{2}\right) .  \label{equa}
\end{equation}%
Introducing the conformal time%
\begin{equation}
\tau =-\int_{t}^{+\infty }\frac{\mathrm{d}t^{\prime }}{a(t^{\prime })},
\label{tau}
\end{equation}%
equation (\ref{equa}) can be converted into the beta function $\beta
_{\alpha }\equiv \mathrm{d}\alpha /\mathrm{d\ln }|\tau |$ of the cosmic RG
flow, which can be easily worked out to arbitrarily high orders in $\alpha $%
. For example, to order $\alpha ^{6}$ we find 
\begin{equation}
\beta _{\alpha }=-2\alpha ^{2}\left[ 1+\frac{5}{6}\alpha +\frac{25}{9}\alpha
^{2}+\frac{383}{27}\alpha ^{3}+\frac{8155}{81}\alpha ^{4}+\frac{72206}{81}%
\alpha ^{5}+\frac{2367907}{243}\alpha ^{6}+\mathcal{O}(\alpha ^{7})\right] .
\label{beta}
\end{equation}

The running coupling $\alpha (x)$ is the solution of%
\begin{equation*}
\ln \frac{\tau }{\tau ^{\prime }}=\int_{\alpha (-\tau ^{\prime })}^{\alpha
(-\tau )}\frac{\mathrm{d}\alpha }{\beta _{\alpha }(\alpha )}.
\end{equation*}%
Throughout this paper we use the notations $\alpha $ for $\alpha (-\tau )$
and $\alpha _{k}$ for $\alpha (1/k)$, where $k$ is just a constant for now.
Later on $k$ will denote the absolute value of the space momentum of the
fluctuations. We have%
\begin{equation*}
\ln (-k\tau )=\int_{\alpha _{k}}^{\alpha }\frac{\mathrm{d}\alpha ^{\prime }}{%
\beta _{\alpha }(\alpha ^{\prime })}.
\end{equation*}%
For example, the leading-log running coupling is%
\begin{equation}
\alpha =\frac{\alpha _{k}}{1+2\alpha _{k}\ln (-k\tau )}.  \label{arun}
\end{equation}%
The expression of the running coupling to the NNLL order can be found in 
\cite{CMBrunning} or appendix \ref{formulas}.

It can be proved \cite{CMBrunning} that the spectra $\mathcal{P}_{T}$ and $%
\mathcal{P}_{\mathcal{R}}$ of the tensor and scalar fluctuations satisfy RG
evolution equations in the superhorizon limit with vanishing anomalous
dimensions. Here we summarize various versions of the equations. Viewing the
spectra as functions of $\tau $ and $\alpha $, they satisfy%
\begin{equation}
\frac{\mathrm{d}\mathcal{P}}{\mathrm{d}\ln |\tau |}=\left( \frac{\partial }{%
\partial \ln |\tau |}+\beta _{\alpha }(\alpha )\frac{\partial }{\partial
\alpha }\right) \mathcal{P}=0.  \label{RG}
\end{equation}%
Viewing them as functions of $\alpha $ and $\alpha _{k}$, the RG equations
imply that the dependence on $\alpha $ actually drops out, 
\begin{equation}
\mathcal{P}=\mathcal{\tilde{P}}(\alpha _{k}),\qquad \frac{\mathrm{d}\mathcal{%
\tilde{P}}(\alpha _{k})}{\mathrm{d}\ln k}=-\beta _{\alpha }(\alpha _{k})%
\frac{\mathrm{d}\mathcal{\tilde{P}}(\alpha _{k})}{\mathrm{d}\alpha _{k}},
\label{noalfa}
\end{equation}%
so the spectra depend on the momentum $k$ only through the running coupling $%
\alpha _{k}$. Finally, viewing the spectra as functions of $k/k_{\ast }$ and 
$\alpha _{\ast }=\alpha (1/k_{\ast })$, where $k_{\ast }$ is the pivot scale
and $\alpha _{\ast }$ is the \textquotedblleft pivot
coupling\textquotedblright , the spectra satisfy%
\begin{equation}
\left( \frac{\partial }{\partial \ln k}+\beta _{\alpha }(\alpha _{\ast })%
\frac{\partial }{\partial \alpha _{\ast }}\right) \mathcal{P}(k/k_{\ast
},\alpha _{\ast })=0.  \label{RGeq}
\end{equation}

As said, the RG techniques allow us to calculate RG improved power spectra.
This means that $\mathcal{P}_{T}$ and $\mathcal{P}_{\mathcal{R}}$ are
expanded in powers of $\alpha _{\ast }$, but the product $\alpha _{\ast }\ln
(k/k_{\ast })$ is considered of order zero and treated exactly.

\section[Limit of heavy fakeon]{Limit $m_{\chi }\rightarrow \infty $ of
infinitely heavy fakeon}

\label{R2}\setcounter{equation}{0}

In this section we derive the running power spectra to the NNLL order in the
limit of infinitely heavy fakeon $m_{\chi }\rightarrow \infty $, which
returns the Starobinsky $R+R^{2}$ theory \cite{starobinsky,vilenkin}. The
action is%
\begin{equation}
S=-\frac{1}{16\pi G}\int \mathrm{d}^{4}x\sqrt{-g}R+\frac{1}{2}\int \mathrm{d}%
^{4}x\sqrt{-g}\left( D_{\mu }\phi D^{\mu }\phi -2V(\phi )\right) ,
\label{staroac}
\end{equation}%
with the potential (\ref{staropote}). We parametrize the metric as 
\begin{eqnarray}
g_{\mu \nu } &=&\text{diag}(1,-a^{2},-a^{2},-a^{2})-2a^{2}\left( u\delta
_{\mu }^{1}\delta _{\nu }^{1}-u\delta _{\mu }^{2}\delta _{\nu }^{2}+v\delta
_{\mu }^{1}\delta _{\nu }^{2}+v\delta _{\mu }^{2}\delta _{\nu }^{1}\right) ,
\notag \\
&&+2\text{diag}(\Phi ,a^{2}\Psi ,a^{2}\Psi ,a^{2}\Psi )-\delta _{\mu
}^{0}\delta _{\nu }^{i}\partial _{i}B-\delta _{\mu }^{i}\delta _{\nu
}^{0}\partial _{i}B.  \label{mets}
\end{eqnarray}%
Without loss of generality, the coordinate dependencies of the graviton
modes $u=u(t,z)$ and $v=v(t,z)$ are chosen to have a space momentum oriented
along the $z$ axis after Fourier transform. As far as the scalar modes are
concerned, we work in the comoving gauge, where the $\phi $ fluctuation $%
\delta \phi $ vanishes and the curvature perturbation $\mathcal{R}$\
coincides with $\Psi $. For reviews that contain details on the
parametrizations of the metric fluctuations and their transformations under
diffeomorphisms, see \cite{baumann,reviews}.

The tensor fluctuations are studied by setting $\Phi =\Psi =B=0$. Denoting
the Fourier transform of $u(t,z)$ with respect to the coordinate $z$ by $u_{%
\mathbf{k}}(t)$, where $\mathbf{k}$ is the space momentum, the quadratic
Lagrangian obtained from (\ref{staroac}) is 
\begin{equation}
(8\pi G)\frac{\mathcal{L}_{\text{t}}}{a^{3}}=\dot{u}_{\mathbf{k}}\dot{u}_{-%
\mathbf{k}}-\frac{k^{2}}{a^{2}}u_{\mathbf{k}}u_{-\mathbf{k}},  \label{lut}
\end{equation}%
plus an identical contribution for $v_{\mathbf{k}}$, where $k=|\mathbf{k}|$.
We often drop the subscripts $\mathbf{k}$ and $-\mathbf{k}$, when no
confusion is expected to arise.

The scalar fluctuations are studied by setting $u=v=0$. After Fourier
transforming the space coordinates to momentum space, (\ref{sqgeq}) gives
the quadratic Lagrangian 
\begin{equation}
(8\pi G)\frac{\mathcal{L}_{\text{s}}}{a^{3}}=-3(\dot{\Psi}+H\Phi )^{2}+4\pi G%
\dot{\phi}^{2}\Phi ^{2}+\frac{k^{2}}{a^{2}}\left[ 2B(\dot{\Psi}+H\Phi )+\Psi
(\Psi -2\Phi )\right] ,  \notag
\end{equation}%
omitting the subscripts $\mathbf{k}$ and $-\mathbf{k}$. Integrating $B$ out,
we obtain $\Phi =-\dot{\Psi}/H$. Inserting this solution back into the
action, we find 
\begin{equation}
(8\pi G)\frac{\mathcal{L}_{\text{s}}}{a^{3}}=3\alpha ^{2}\left( \dot{\Psi}%
^{2}-\frac{k^{2}}{a^{2}}\Psi ^{2}\right) .  \label{ls}
\end{equation}

Defining%
\begin{equation}
w=au\sqrt{\frac{k}{4\pi G}},\qquad w=\alpha a\Psi \sqrt{\frac{3k}{4\pi G}},
\label{w}
\end{equation}%
for tensors and scalars, respectively, and switching to the variable $\eta
=-k\tau $, the Lagrangians (\ref{lut}) and (\ref{ls})\ give the actions%
\begin{equation}
S_{\text{t,s}}=\frac{1}{2}\int \mathrm{d}\eta \left[ w^{\prime \hspace{0.01in%
}2}-w^{2}+(2+\sigma _{\text{t,s}})\frac{w^{2}}{\eta ^{2}}\right] ,
\label{sred}
\end{equation}%
where the prime denotes the derivative with respect to $\eta $ and%
\begin{equation}
\sigma _{\text{t}}=9\alpha ^{2}+48\alpha ^{3}+364\alpha ^{4}+\mathcal{O}%
(\alpha ^{5}),\qquad \sigma _{\text{s}}=6\alpha +22\alpha ^{2}+\frac{280}{3}%
\alpha ^{3}+\mathcal{O}(\alpha ^{4}).  \label{sigmas}
\end{equation}

We quantize (\ref{sred}) as usual, by introducing the operator 
\begin{equation*}
\hat{w}_{\mathbf{k}}(\eta )=w_{\mathbf{k}}(\eta )\hat{a}_{\mathbf{k}}+w_{-%
\mathbf{k}}^{\ast }(\eta )\hat{a}_{-\mathbf{k}}^{\dagger },
\end{equation*}%
where $\hat{a}_{\mathbf{k}}^{\dagger }$ and $\hat{a}_{\mathbf{k}}$ are
creation and annihilation operators satisfying $[\hat{a}_{\mathbf{k}},\hat{a}%
_{\mathbf{k}^{\prime }}^{\dagger }]=(2\pi )^{3}\delta ^{(3)}(\mathbf{k}-%
\mathbf{k}^{\prime })$. Summing over the tensor polarizations $u$ and $v$
and recalling that $\mathcal{R}=\Psi $, the power spectra $\mathcal{P}_{T}$
and $\mathcal{P}_{\mathcal{R}}$ of the tensor and scalar fluctuations are
defined by the two-point functions%
\begin{eqnarray}
\langle \hat{u}_{\mathbf{k}}(\tau )\hat{u}_{\mathbf{k}^{\prime }}(\tau
)\rangle  &=&(2\pi )^{3}\delta ^{(3)}(\mathbf{k}+\mathbf{k}^{\prime })\frac{%
\pi ^{2}}{8k^{3}}\mathcal{P}_{T},\qquad \mathcal{P}_{T}=\frac{8k^{3}}{\pi
^{2}}|u_{\mathbf{k}}|^{2},  \label{pt} \\
\langle \mathcal{R}_{\mathbf{k}}(\tau )\mathcal{R}_{\mathbf{k}^{\prime
}}(\tau )\rangle  &=&(2\pi )^{3}\delta ^{(3)}(\mathbf{k}+\mathbf{k}^{\prime
})\frac{2\pi ^{2}}{k^{3}}\mathcal{P}_{\mathcal{R}},\qquad \mathcal{P}_{%
\mathcal{R}}=\frac{k^{3}}{2\pi ^{2}}|\Psi _{\mathbf{k}}|^{2}.  \label{pR}
\end{eqnarray}

The calculations of $\mathcal{P}_{T}$ and $\mathcal{P}_{\mathcal{R}}$ are
divided in two steps. In the first step we isolate the time dependence,
which disappears in the superhorizon limit by the RG equation (\ref{RG}). In
the second step we work out the overall constants (which are functions of $%
\alpha _{k}$).

We start from the Mukhanov-Sasaki equation derived from the action (\ref%
{sred}), which reads%
\begin{equation}
w^{\prime \prime }+w-2\frac{w}{\eta ^{2}}=\sigma _{\text{t,s}}\frac{w}{\eta
^{2}}  \label{mukhagen}
\end{equation}%
and must be solved with the Bunch-Davies vacuum condition%
\begin{equation}
w(\eta )\sim \frac{\mathrm{e}^{i\eta }}{\sqrt{2}}\qquad \text{for large }%
\eta .  \label{bunch}
\end{equation}

To study the time dependence in the superhorizon limit, it is convenient to
decompose $\eta w(\eta )$ as the sum of a power series $Q(\ln \eta )$ in $%
\ln \eta $ plus a power series $W(\eta )$ in $\eta $ and $\ln \eta $, such
that $W(\eta )\rightarrow 0$ term-by-term for $\eta \rightarrow 0$:%
\begin{equation}
\eta w=Q(\ln \eta )+W(\eta ).  \label{decompo}
\end{equation}%
It was shown in ref. \cite{CMBrunning} that the $w$ equation (\ref{mukhagen}%
) leads to the $Q$ equation 
\begin{equation}
\frac{\mathrm{d}Q}{\mathrm{d}\ln \eta }=-\frac{\sigma }{3}Q-\frac{1}{3}%
\sum_{n=1}^{\infty }3^{-n}\frac{\mathrm{d}^{n}(\sigma Q)}{\mathrm{d}\ln
^{n}\eta },  \label{Pw}
\end{equation}%
where $\sigma $ stands for either $\sigma _{\text{t}}$ or $\sigma _{\text{s}%
} $ and the higher-derivative terms on the right-hand side have to be
handled perturbatively in $\alpha _{k}$. We can also view $Q(\ln \eta )$ as
a function $\tilde{Q}(\alpha ,\alpha _{k})$ of $\alpha $ and $\alpha _{k}$,
satisfying%
\begin{equation}
\beta _{\alpha }\frac{\partial \tilde{Q}}{\partial \alpha }=-\frac{\sigma 
\tilde{Q}}{3}-\frac{1}{3}\sum_{n=1}^{\infty }3^{-n}\left( \beta _{\alpha }%
\frac{\partial }{\partial \alpha }\right) ^{n}(\sigma \tilde{Q}).
\label{Pwt}
\end{equation}

Since the RG equations hold only in the superhorizon limit, where the
Bunch-Davies vacuum condition is ineffective, equation (\ref{Pw}) does not
give the normalization of the spectrum. The initial condition $Q(0)$, which
remains arbitrary at this stage, is determined in the second step of the
calculation, by working directly on $w$ and the equation (\ref{mukhagen}).

Formulas (\ref{sigmas}) are enough to work out the NNLL corrections. For
that purpose, we can truncate (\ref{Pwt}) to%
\begin{equation}
\beta _{\alpha }\frac{\partial \tilde{Q}}{\partial \alpha }=-\frac{\sigma 
\tilde{Q}}{3}-\frac{\beta _{\alpha }}{9}\frac{\partial (\sigma \tilde{Q})}{%
\partial \alpha }-\frac{\beta _{\alpha }}{27}\frac{\partial }{\partial
\alpha }\left( \beta _{\alpha }\frac{\partial (\sigma \tilde{Q})}{\partial
\alpha }\right)  \label{trun}
\end{equation}%
and the beta function to 
\begin{equation*}
\beta _{\alpha }=-2\alpha ^{2}-\frac{5}{3}\alpha ^{3}-\frac{50}{9}\alpha
^{4}+\mathcal{O}(\alpha ^{5}),
\end{equation*}

We solve (\ref{trun}) by searching for solutions of the form%
\begin{equation*}
\tilde{Q}_{\text{t}}(\alpha ,\alpha _{k})=\tilde{Q}_{\text{t}}(\alpha _{k})%
\frac{1+\sum_{j=1}^{\infty }c_{j}\alpha ^{j}}{1+\sum_{j=1}^{\infty
}c_{j}\alpha _{k}^{j}},\qquad \tilde{Q}_{\text{s}}(\alpha ,\alpha _{k})=%
\tilde{Q}_{\text{s}}(\alpha _{k})\frac{\alpha }{\alpha _{k}}\frac{%
1+\sum_{j=1}^{\infty }c_{j}^{\prime }\alpha ^{j}}{1+\sum_{j=1}^{\infty
}c_{j}^{\prime }\alpha _{k}^{j}},
\end{equation*}%
for tensors and scalars, respectively, where $c_{j}$ and $c_{j}^{\prime }$
are coefficients to be determined. Inserting these expressions into (\ref%
{trun}), we easily find%
\begin{eqnarray}
\tilde{Q}_{\text{t}}(\alpha ,\alpha _{k}) &=&\tilde{Q}_{\text{t}}(\alpha
_{k})\frac{2+3\alpha +7\alpha ^{2}+\frac{199}{6}\alpha ^{3}+\mathcal{O}%
(\alpha ^{4})}{2+3\alpha _{k}+7\alpha _{k}^{2}+\frac{199}{6}\alpha _{k}^{3}+%
\mathcal{O}(\alpha _{k}^{4})},  \notag \\
\tilde{Q}_{\text{s}}(\alpha ,\alpha _{k}) &=&\tilde{Q}_{\text{s}}(\alpha
_{k})\frac{\alpha }{\alpha _{k}}\frac{2+3\alpha +7\alpha ^{2}+\mathcal{O}%
(\alpha ^{3})}{2+3\alpha _{k}+7\alpha _{k}^{2}+\mathcal{O}(\alpha _{k}^{3})}.
\label{Qtilde}
\end{eqnarray}%
As said, the constants $\tilde{Q}_{\text{t,s}}(\alpha _{k})=\tilde{Q}_{\text{%
t,s}}(\alpha _{k},\alpha _{k})=Q_{\text{t,s}}(0)$ are determined separately.

Formula (\ref{decompo}) gives $\eta w\sim Q(\ln \eta )$ in the superhorizon
limit, while formulas (\ref{w}) give $u$ and $\Psi $. Together with formulas
(\ref{acca}) and (\ref{aHtau}) to order $\alpha ^{3}$, it is easy to show
that the spectra (\ref{pt}) and (\ref{pR}) are equal to 
\begin{eqnarray}
\mathcal{P}_{T}(k) &=&\frac{8Gm_{\phi }^{2}}{\pi }|\tilde{Q}_{\text{t}%
}(\alpha _{k})|^{2}\left[ 1-3\alpha _{k}-\frac{\alpha _{k}^{2}}{4}-\frac{91}{%
6}\alpha _{k}^{3}+\mathcal{O}(\alpha _{k}^{4})\right] ,  \notag \\
\mathcal{P}_{\mathcal{R}}(k) &=&\frac{Gm_{\phi }^{2}}{6\pi \alpha _{k}^{2}}|%
\tilde{Q}_{\text{s}}(\alpha _{k})|^{2}\left[ 1-3\alpha _{k}-\frac{\alpha
_{k}^{2}}{4}+\mathcal{O}(\alpha _{k}^{3})\right] ,  \label{ptprQ}
\end{eqnarray}%
respectively, in the superhorizon limit. We see that the $\alpha $
dependence disappears, in agreement with the RG equations (\ref{RG}) and (%
\ref{noalfa}). The dependence on $k$ is encoded into the running coupling $%
\alpha _{k}=\alpha (1/k)$. The similarity between the two square brackets of
(\ref{ptprQ}) will not survive the extension to $m_{\chi }<\infty $.

Now we calculate the integration constants $\tilde{Q}_{\text{t,s}}(\alpha
_{k})=Q_{\text{t,s}}(0)$. We expand $w$ in powers of $\alpha _{k}$ by
writing 
\begin{equation}
w(\eta )=w_{0}(\eta )+\sum_{n=1}^{\infty }\alpha _{k}^{n}w_{n}(\eta ),
\label{wetasstaro}
\end{equation}%
and insert the expansion into (\ref{mukhagen}). We obtain equations of the
form%
\begin{equation}
w_{n}^{\prime \prime }+w_{n}-\frac{2w_{n}}{\eta ^{2}}=\frac{g_{n}(\eta )}{%
\eta ^{2}},  \label{weqsstaro}
\end{equation}%
where the functions $g_{n}$ are determined recursively from $w_{m}$, $m<n$.
The Bunch-Davies vacuum condition (\ref{bunch}) gives $w_{0}(\eta )\sim 
\mathrm{e}^{i\eta }/\sqrt{2}$ and $w_{m}(\eta )\rightarrow 0$, $m>0$, for
large $\eta .$

We need the solutions $w_{i}$ with $i=0,1,2,3$ for tensors and $i=0,1,2$ for
scalars, which are reported in the appendix. The behaviors of $w_{i}\left(
\eta \right) $ in the superhorizon limit $\eta \sim 0$, also given in
appendix \ref{formulas}, allow us to derive $w(\eta )$ for $\eta \sim 0$,
once the solutions $w_{i}$ are inserted into (\ref{wetasstaro}). Formula (%
\ref{decompo}) tells us that $\eta w_{\text{t,s}}(\eta )\sim Q_{\text{t,s}%
}(\ln \eta )$ in the same limit. From that, we can read $Q_{\text{t,s}}(\ln
\eta )$ and hence $Q_{\text{t,s}}(0)$. After these calculations, we find%
\begin{eqnarray}
\tilde{Q}_{\text{t}}(\alpha _{k}) &=&\frac{i}{\sqrt{2}}\left[ 1+3(2-\tilde{%
\gamma}_{M})\alpha _{k}^{2}+\left( 12\tilde{\gamma}_{M}-6\tilde{\gamma}%
_{M}^{2}-\pi ^{2}\right) \alpha _{k}^{3}+\mathcal{O}(\alpha _{k}^{4})\right]
,  \notag \\
\tilde{Q}_{\text{s}}(\alpha _{k}) &=&\frac{i}{\sqrt{2}}\left[ 1+2(2-\tilde{%
\gamma}_{M})\alpha _{k}+\frac{2}{3}(2-7\tilde{\gamma}_{M}+\pi ^{2})\alpha
_{k}^{2}+\mathcal{O}(\alpha _{k}^{3})\right] ,  \label{Q0staro}
\end{eqnarray}%
where $\tilde{\gamma}_{M}=\gamma _{M}-(i\pi /2)$, $\gamma _{M}=\gamma
_{E}+\ln 2$, $\gamma _{E}$ being the Euler-Mascheroni constant.

A further check of the RG equations can be made by noting that at $\ln \eta
\neq 0$, the expressions of $Q_{\text{t}}(\ln \eta )$ and $Q_{\text{s}}(\ln
\eta )$ just obtained must coincide with (\ref{Qtilde}) -- equipped with (%
\ref{Q0staro}) -- up to the RG improvements. This means that, once $\alpha $
is replaced with the running coupling given in the appendix, which is a
function of $\ln \eta $ and $\alpha _{k}$, $Q_{\text{t}}(\ln \eta )$ must
agree with $\tilde{Q}_{\text{t}}(\alpha ,\alpha _{k})$ to order $\alpha
_{k}^{3}$, while $Q_{\text{s}}(\ln \eta )$ must agree with $\tilde{Q}_{\text{%
s}}(\alpha ,\alpha _{k})$ to order $\alpha _{k}^{2}$. It is easy to verify
that both properties are satisfied. We recall that the enhancement contained
in (\ref{Qtilde}) is the RG improvement, i.e. the resummation of the leading
and subleading logs to order $\alpha ^{3}$.

Finally, inserting the normalizations (\ref{Q0staro}) into (\ref{ptprQ}) the
spectra are%
\begin{eqnarray}
\mathcal{P}_{T}(k) &=&\frac{4Gm_{\phi }^{2}}{\pi }\left[ 1-3\alpha
_{k}+(47-24\gamma _{M})\frac{\alpha _{k}^{2}}{4}-\left( \frac{307}{6}%
+12\gamma _{M}^{2}-42\gamma _{M}-\pi ^{2}\right) \alpha _{k}^{3}+\!\mathcal{O%
}(\alpha _{k}^{4})\right] \!\!,  \notag \\
\mathcal{P}_{\mathcal{R}}(k) &=&\frac{Gm_{\phi }^{2}}{12\pi \alpha _{k}^{2}}%
\left[ 1+(5-4\gamma _{M})\alpha _{k}-\frac{67}{12}\alpha _{k}^{2}+(12\gamma
_{M}^{2}-40\gamma _{M}+7\pi ^{2})\frac{\alpha _{k}^{2}}{3}+\mathcal{O}%
(\alpha _{k}^{3})\right] \!\!.\!\!  \label{ptpr}
\end{eqnarray}%
The tilts and the running coefficients can be found by differentiating with
respect to $\ln k$, using the RG\ equations and the expression of the beta
function. We postpone this part and proceed to compute the spectra of
quantum gravity.

\section{Quantum gravity: tensor fluctuations}

\label{tensorQG}

In this section we compute the spectrum of the tensor fluctuations to the
NNLL\ order in quantum gravity.

Parametrizing the metric as (\ref{mets}) with $\Phi =\Psi =B=0$, the
quadratic Lagrangian obtained from (\ref{sqgeq}) is 
\begin{equation}
(8\pi G)\frac{\mathcal{L}_{\text{t}}}{a^{3}}=\dot{u}^{2}-\frac{k^{2}}{a^{2}}%
u^{2}-\frac{1}{m_{\chi }^{2}}\left[ \ddot{u}^{2}-2\left( H^{2}-\frac{3}{2}%
\alpha ^{2}H^{2}+\frac{k^{2}}{a^{2}}\right) \dot{u}^{2}+\frac{k^{4}}{a^{4}}%
u^{2}\right] ,  \label{lt}
\end{equation}%
plus an identical contribution for $v$. The calculation proceeds as follows.
First, we eliminate the higher derivatives from $\mathcal{L}_{\text{t}}$ by
introducing an extra field. Then, we diagonalize the new Lagrangian and
perform the fakeon projection. As expected, we obtain a projected Lagrangian
that depends on just one field and has no higher derivatives. Third, we
apply a number of field redefinitions and time reparametrizations to cast
the action into the standard form (\ref{sred}). Then we solve the $w$
equation with the Bunch-Davies vacuum condition. Finally, we undo all the
transformations and work out the $u$ two-point function and the tensor
spectrum in the superhorizon limit.

To calculate the spectra to the NNLL order we need the projected Lagrangian
to order $\alpha ^{3}$. We also plan to make an independent check the RG
evolution equations. For that purpose, the $k$-independent corrections to
the Lagrangian are needed to order $\alpha ^{4}$.

The higher derivatives of $\mathcal{L}_{\text{t}}$ can be eliminated with
the procedure outlined in \cite{ABP}. Specifically, we add an auxiliary
field $U$ and consider the extended Lagrangian%
\begin{equation}
\mathcal{L}_{\text{t}}^{\prime }=\mathcal{L}_{\text{t}}+\frac{a^{3}}{8\pi
Gm_{\chi }^{2}}\left[ m_{\chi }^{2}\sqrt{\gamma }U-\ddot{u}-3H\left( 1-\frac{%
4\alpha ^{2}H^{2}}{m_{\chi }^{2}\gamma }\right) \dot{u}-fu\right] ^{2},
\label{ltp}
\end{equation}%
where\footnote{%
Note some rearrangements with respect to the parametrizations of \cite{ABP}
and \cite{CMBrunning}, for a better inclusion of the extra order we need.} 
\begin{equation}
f=m_{\chi }^{2}\gamma +\frac{k^{2}}{a^{2}}+\frac{\alpha ^{2}H^{2}}{m_{\chi
}^{2}\gamma }\left( 3m_{\chi }^{2}-12H^{2}+24\alpha H^{2}-\frac{2\alpha
^{2}H^{2}(17m_{\chi }^{2}-38H^{2})}{m_{\chi }^{2}\gamma }\right) \qquad 
\notag
\end{equation}%
and 
\begin{equation}
\gamma =1+2\frac{H^{2}}{m_{\chi }^{2}}.  \label{gam}
\end{equation}%
It is straightforward to show that $\mathcal{L}_{\text{t}}^{\prime }$ is
equivalent to $\mathcal{L}_{\text{t}}$ by replacing $U$, which appears
algebraically, with the solution of its own field equation. As said, the
reason why we keep the order $\alpha ^{4}$ is that we need it to check the
RG equation to order $\alpha ^{3}$. If we are happy with just using the RG
equation, which we know to hold since it was proved on general grounds in 
\cite{CMBrunning}, the Lagrangian to order $\alpha ^{3}$ is sufficient for
our purposes.

We diagonalize $\mathcal{L}_{\text{t}}^{\prime }$ in the de Sitter limit by
making the field redefinition 
\begin{equation}
u=\frac{U+V}{\sqrt{\gamma }},  \label{uU}
\end{equation}%
where $V$ is a new field. To the order we need, the ``de-Sitter-diagonal''\ $%
\mathcal{L}_{\text{t}}^{\prime }$ reads%
\begin{eqnarray}
(8\pi G)\frac{\mathcal{L}_{\text{t}}^{\prime }}{a^{3}} &=&\dot{U}^{2}-\frac{%
hk^{2}}{a^{2}}U^{2}-\frac{9}{8}m_{\phi }^{2}\xi \zeta \alpha ^{2}\left[ 1-%
\frac{\zeta \alpha }{6}(40-7\xi )+\frac{\zeta ^{2}\alpha ^{2}}{144}%
(2800-3806\xi -497\xi ^{2})\right] U^{2}  \notag \\
&&-\dot{V}^{2}+\left[ m_{\chi }^{2}+\frac{m_{\phi }^{2}}{2}(1-3\alpha )+%
\frac{k^{2}}{a^{2}}\right] V^{2}+\frac{3m_{\phi }^{2}\zeta \alpha ^{2}}{2}%
\left( 1-\xi +\frac{4\xi \zeta k^{2}}{m_{\phi }^{2}a^{2}}\right) UV  \notag
\\
&&-\frac{3m_{\phi }^{2}\zeta ^{2}\alpha ^{3}}{4}\left[ 6-22\xi +\xi
^{2}+(6-7\xi -2\xi ^{2})\frac{4\xi \zeta k^{2}}{m_{\phi }^{2}a^{2}}\right]
UV,  \label{ltg}
\end{eqnarray}%
where%
\begin{equation}
\xi =\frac{m_{\phi }^{2}}{m_{\chi }^{2}},\qquad \zeta =\left( 1+\frac{\xi }{2%
}\right) ^{-1},\qquad h=1-3\xi \zeta ^{2}\alpha ^{2}+\frac{3\xi \zeta
^{3}\alpha ^{3}}{2}(6-7\xi -2\xi ^{2})+\mathcal{O}(\alpha ^{4}).
\label{cigacca}
\end{equation}%
It will be clear in a moment that it is enough to keep the terms
proportional to $V^{2}$ to order $\alpha $ and the terms proportional to $UV$%
, as well as those proportional to $k^{2}U^{2}$, to order $\alpha ^{3}$, as
done in (\ref{ltg}) and (\ref{cigacca}).

\subsection{The fakeon projection}

The fakeon projection amounts to integrating $V$ out by replacing it with a
particular solution $V(U)$ of its own field equations, defined by the fakeon
Green function \cite{ABP}\footnote{%
See also \cite{classicization,FLRW}.}. From the expression of (\ref{ltg}) we
see that the projection equates $V$ to something of order $\alpha ^{2}$.
Once the solution is inserted back into (\ref{ltg}), the second and third
lines of (\ref{ltg}) turn out to be $\mathcal{O}(\alpha ^{4})$. This means
that to the order $\alpha ^{3}$ included, which is enough for our present
purposes, the projected $U$ action is unaffected by $V(U)$. Specifically, it
is given by the first line of (\ref{ltg}) and can be solved with the
standard Bunch-Davies vacuum condition\footnote{%
At higher orders, instead, the projected $U$ action receives contributions
from $V(U)$ and depends on $k$ in a nontrivial way. Then the Bunch-Davies
vacuum condition for large $k$ needs to be reconsidered and possibly
replaced by a different condition.}.

This does not mean we can forget about $V$ altogether, however. Indeed,
formula (\ref{uU}) tells us that we need $V(U)$ to calculate the spectrum.
Since we want the spectrum to the NNLL\ order, we must compute $V(U)$ to
order $\alpha ^{3}$, for which (\ref{ltg}) is enough. A lot of effort can be
saved by noting that it is sufficient to obtain $V(U)$ in the superhorizon
limit $k/(m_{\phi }a)\rightarrow 0$. Moreover, we can drop several
contributions by making use of the $U$ field equations.

With this in mind, the projected $V$ can be easily found by inserting the
ansatz%
\begin{equation*}
V=\alpha ^{2}\left( v_{1}+v_{2}\alpha \right) U+v_{3}\alpha ^{3}\dot{U}+%
\mathcal{O}(\alpha ^{4})
\end{equation*}%
into the $V$ field equation derived from\ (\ref{ltg}), where $v_{i}$, $%
i=1,2,3$, are constants. Then we take the superhorizon limit by dropping the 
$k$-dependent terms. Third, we use the $U$ field equation to turn the
higher-derivatives $\ddot{U}$ and $U^{(3)}$ into corrections proportional to 
$U$ and $\dot{U}$. At the end, we obtain a linear combination of $\alpha
^{2}U$, $\alpha ^{3}U$ and $\alpha ^{3}\dot{U}$. Equating the coefficients
of such three terms to zero, we determine the constants $v_{i}$. The result
gives%
\begin{equation}
V=-\frac{3\xi \zeta ^{2}\alpha ^{2}}{4}\left[ 1-\xi -(6-19\xi -2\xi ^{2})%
\frac{\zeta \alpha }{2}\right] U+\frac{3(1-\xi )\xi ^{2}\zeta ^{3}\alpha ^{3}%
}{m_{\phi }}\dot{U}+\mathcal{O}(\alpha ^{4}).  \label{v}
\end{equation}

\subsection[w action and RG equation]{$w$ action and RG\ equation}

We first calculate the spectrum to the NNLL order by means of the RG
equation. In a second moment, we check the RG equation and the spectrum
independently. As said, the tensor analogue of the Mukhanov-Sasaki action to
order $\alpha ^{3}$ is given by the first line of (\ref{ltg}). We define%
\begin{equation}
w=\frac{aU\sqrt{k}}{\sqrt{4\pi G}},\qquad \sigma _{\text{t}}=9\zeta \alpha
^{2}+\frac{3\zeta ^{2}\alpha ^{3}}{2}(32+43\xi )+\zeta ^{3}\alpha ^{4}F_{%
\text{t}}(\xi )+\mathcal{O}(\alpha ^{5}),  \label{wsi}
\end{equation}%
where the function $F_{\text{t}}(\xi )$ parametrizes the order $\alpha ^{4}$
of $\sigma _{\text{t}}$, for the moment unknown. Using (\ref{acca}) and (\ref%
{aHtau}) and switching to conformal time, the projected $w$ action to the
order we need can be cast into the form%
\begin{equation}
S_{\text{t}}^{\text{prj}}=\frac{1}{2}\int \mathrm{d}\eta \left( w^{\prime 
\hspace{0.01in}2}-hw^{2}+2\frac{w^{2}}{\eta ^{2}}+\sigma _{\text{t}}\frac{%
w^{2}}{\eta ^{2}}\right) .  \label{sredF}
\end{equation}%
The unusual feature of this expression is the function $h$ in front of $w^{2}
$, which is a sort of running squared mass.

Having parametrized the order $\alpha ^{4}$ of $\sigma _{\text{t}}$ as shown
in (\ref{wsi}), we first determine $F_{\text{t}}(\xi )$ by means of the RG
equation and then work out the spectrum to the NNLL order. Later on, in
subsection \ref{checks}, we make an independent check of the RG equation by
calculating $\sigma _{\text{t}}$ directly to order $\alpha ^{4}$, with the
help of an expansion in powers of $\xi $, and rederive the spectrum without
using the RG equation.

It is important to recall that the RG equation holds in the superhorizon
limit, where the Bunch-Davies condition is unnecessary and we can ignore the
term $-hw^{2}$ of (\ref{sredF}). In particular, once we decompose $w$ as
shown in (\ref{decompo}), the equations (\ref{Pw}) and (\ref{Pwt}) still
hold, so we can calculate the function $Q_{\text{t}}(\ln \eta )=\tilde{Q}_{%
\text{t}}(\alpha ,\alpha _{k})$ by solving (\ref{trun}) with the $\sigma _{%
\text{t}}$ of (\ref{wsi}) and repeating the steps from (\ref{decompo}) to (%
\ref{Qtilde}). We find%
\begin{equation}
\tilde{Q}_{\text{t}}(\alpha ,\alpha _{k})=\frac{J_{\text{t}}(\alpha )}{J_{%
\text{t}}(\alpha _{k})}\tilde{Q}_{\text{t}}(\alpha _{k}),  \label{Qtt}
\end{equation}%
with 
\begin{equation}
J_{\text{t}}(\alpha )=1+\frac{3\zeta \alpha }{2}+\frac{56+73\xi }{16}\zeta
^{2}\alpha ^{2}+\left( F_{\text{t}}(\xi )-\frac{131}{2}-\frac{2029\xi }{16}-%
\frac{1441\xi ^{2}}{16}\right) \frac{\zeta ^{3}\alpha ^{3}}{18}+\mathcal{O}%
(\alpha ^{4}).  \notag
\end{equation}

At this point, we go back to $u$ as follows. We first use (\ref{decompo}) to
compute $w$ in the superhorizon limit, where $\eta w\sim Q(\ln \eta )=\tilde{%
Q}(\alpha ,\alpha _{k})$. Then we use the first formula of (\ref{wsi}) to
compute $U$, (\ref{v}) to compute $V$ and finally (\ref{uU}) to\ compute $u$.

The RG equation (\ref{RG}) implies that $u$ is time independent in the
superhorizon limit. In the parametrization we are using, which is the one of
(\ref{noalfa}), $u$ can depend on $\alpha _{k}$, but not on $\alpha $. It is
easy to check that, indeed, no $\alpha $ dependence appears to order $\alpha
^{3}$. As far as the contributions proportional to $\alpha ^{4}$ are
concerned, they disappear by setting%
\begin{equation}
F_{\text{t}}(\xi )=364+\frac{4037}{8}\xi +\frac{6145}{16}\xi ^{2}+\frac{81}{2%
}\xi ^{3}.  \label{ptx}
\end{equation}%
This is how the RG\ equation fixes $F_{\text{t}}(\xi )$. Now we are ready to
use formula (\ref{pt}), which gives%
\begin{eqnarray}
\mathcal{P}_{T}(k)=\frac{8m_{\phi }^{2}\zeta G}{\pi }|\tilde{Q}_{\text{t}%
}(\alpha _{k})|^{2} &&\left[ 1-3\zeta \alpha _{k}-\frac{\zeta ^{2}\alpha
_{k}^{2}}{8}(2+73\xi )\right.  \notag \\
&&\left. -\frac{\zeta ^{3}\alpha _{k}^{3}}{12}(182+11\xi +392\xi ^{2}+54\xi
^{3})+\mathcal{O}(\alpha _{k}^{4})\right] .  \label{ptQ}
\end{eqnarray}

\subsection[w action and Bunch-Davies vacuum condition]{$w$ action and Bunch-Davies vacuum condition}

For the purpose of calculating $\tilde{Q}_{\text{t}}(\alpha _{k})$, it is
important to deal with the running squared mass encoded in the term $-hw^{2}$
of (\ref{sredF}). Following \cite{CMBrunning}, we change variables from $%
\eta $ to $\tilde{\eta}(\eta )$, such that $\tilde{\eta}^{\prime }(\eta )=%
\sqrt{h(\eta )}$, $\tilde{\eta}(0)=0$, and rewrite (\ref{sredF}) in the more
standard form%
\begin{equation}
\tilde{S}_{\text{t}}^{\text{prj}}=\frac{1}{2}\int \mathrm{d}\tilde{\eta}%
\left( \tilde{w}^{\prime \hspace{0.01in}2}-\tilde{w}^{2}+\frac{2\tilde{w}^{2}%
}{\tilde{\eta}^{2}}+\tilde{\sigma}_{\text{t}}\frac{\tilde{w}^{2}}{\tilde{\eta%
}^{2}}\right) ,  \label{s2}
\end{equation}%
where%
\begin{equation}
\tilde{w}(\tilde{\eta}(\eta ))=h(\eta )^{1/4}w(\eta ),\qquad \tilde{\sigma}_{%
\text{t}}=\frac{\tilde{\eta}^{2}(\sigma _{\text{t}}+2)}{\eta ^{2}h}+\frac{%
\tilde{\eta}^{2}}{16h^{3}}\left( 4hh^{\prime \prime }-5h^{\prime \hspace{%
0.01in}2}\right) -2.  \label{wtilde}
\end{equation}

The Bunch-Davies vacuum condition for (\ref{s2}) is straightforward and reads%
\begin{equation}
\tilde{w}(\tilde{\eta})\simeq \frac{\mathrm{e}^{i\tilde{\eta}}}{\sqrt{2}}%
\text{\qquad for }\tilde{\eta}\rightarrow \infty ,  \label{bdt}
\end{equation}%
as usual. Now we prove that it implies 
\begin{equation}
\tilde{w}(\tilde{\eta})=\frac{(\tilde{\eta}+i)\mathrm{e}^{i\tilde{\eta}}}{%
\sqrt{2}\tilde{\eta}}+\alpha _{k}^{2}\Delta \tilde{w}(\tilde{\eta}),\qquad
\lim_{\tilde{\eta}\rightarrow \infty }\Delta \tilde{w}(\tilde{\eta})=0.
\label{wti}
\end{equation}%
Using (\ref{cigacca}), (\ref{wsi}) and the expression\ (\ref{runca}) of the
running coupling $\alpha $, $\tilde{\sigma}_{\text{t}}$ turns out to be $%
\mathcal{O}\left( \alpha _{k}^{2}\right) $. The solution of the $\tilde{w}$
equation of motion derived from (\ref{s2}) obviously agrees with (\ref{wti})
for $\alpha _{k}=0$, by (\ref{bdt}). This means that we can parametrize $%
\tilde{w}(\tilde{\eta})$ as shown in the first equation of (\ref{wti}), for
a suitable $\Delta \tilde{w}(\tilde{\eta})$. Moreover, the asymptotic
behavior (\ref{bdt}) must hold for every $\alpha _{k}$. This implies that $%
\Delta \tilde{w}(\tilde{\eta})$ must tend to zero for $\tilde{\eta}%
\rightarrow \infty $.

Using (\ref{cigacca}), we find%
\begin{equation*}
\tilde{\eta}=\eta \left[ 1-\frac{3\xi \zeta ^{2}\alpha _{k}^{2}}{2}%
(1-4\alpha _{k}\ln \eta )-\frac{3\xi \zeta ^{3}\alpha _{k}^{3}}{4}(2+11\xi
+2\xi ^{2})+\mathcal{O}(\alpha _{k}^{4})\right] .
\end{equation*}%
Then, $\alpha _{k}^{2}\Delta \tilde{w}(\tilde{\eta})=\alpha _{k}^{2}\Delta 
\tilde{w}(\eta )+\mathcal{O}(\alpha _{k}^{4})$, which allows us to conclude,
using the first formula of (\ref{wtilde}), that $w(\eta )$ has the form%
\begin{equation*}
w(\eta )=\frac{(\tilde{\eta}(\eta )+i)\mathrm{e}^{i\tilde{\eta}(\eta )}}{%
\sqrt{2}\tilde{\eta}(\eta )h(\eta )^{1/4}}+\alpha _{k}^{2}\frac{\Delta 
\tilde{w}(\eta )}{h(\eta )^{1/4}}+\mathcal{O}(\alpha _{k}^{4}),\qquad
\lim_{\eta \rightarrow \infty }\frac{\Delta \tilde{w}(\eta )}{h(\eta )^{1/4}}%
=0.
\end{equation*}%
The Bunch-Davies conditions for $w(\eta )$ to order $\alpha _{k}^{3}$ can be
read from the first term on the right-hand side of the first equation, no
correction coming from the rest. Referring to the expansion (\ref{wetasstaro}%
), we obtain $w_{1}(\eta )=0$ and, for $\eta $ large,$\qquad $%
\begin{eqnarray}
w_{0}(\eta ) &\simeq &\frac{\mathrm{e}^{i\eta }}{\sqrt{2}},\qquad w_{2}(\eta
)\simeq \frac{3\xi \zeta ^{2}\mathrm{e}^{i\eta }(3-2i\eta )}{4\sqrt{2}}, 
\notag \\
w_{3}(\eta ) &\simeq &-\frac{3\xi \zeta ^{2}\mathrm{e}^{i\eta }}{\sqrt{2}}%
(3-2i\eta )\ln \eta -\frac{3\xi \zeta ^{3}\mathrm{e}^{i\eta }}{8\sqrt{2}}%
[2-29\xi -6\xi ^{2}+2i\eta (2+11\xi +2\xi ^{2})].  \label{bdtt}
\end{eqnarray}

\subsection{The spectrum}

Inserting the expansion (\ref{wetasstaro}) into the equation of motion
derived from the action (\ref{sredF}), we obtain the $w_{n}$ equations%
\begin{equation}
w_{n}^{\prime \prime }+w_{n}-2\frac{w_{n}}{\eta ^{2}}=\sum_{j=0}^{n-2}\sigma
_{j}\frac{w_{n-2-j}}{\eta ^{2}}-\sum_{j=0}^{n-2}h_{j}w_{n-2-j},  \label{weq}
\end{equation}%
where%
\begin{equation*}
h=1+\alpha _{k}^{2}\sum_{j=0}^{\infty }h_{j}\alpha _{k}^{j},\qquad \sigma _{%
\text{t}}=\alpha _{k}^{2}\sum_{j=0}^{\infty }\sigma _{j}\alpha _{k}^{j},
\end{equation*}%
$h_{j}$ and $\sigma _{j}$ being functions of $\eta $. The asymptotic
conditions (\ref{bdtt}) determine the solutions.

Although (\ref{weq}) and (\ref{bdtt}) are somewhat different form the
equations and conditions met in the previous section, they can be solved
with similar techniques. The results involve the same types of functions,
listed in appendix \ref{formulas}. Formulas (\ref{gm}) contain the relevant
right-hand sides of (\ref{weq}), while formulas (\ref{wm}) contain the
solutions. Studying the $\eta \sim 0$ limit of $\eta w(\eta )$ with the help
of the asymptotic behaviors (\ref{beha}), we extract $Q_{\text{t}}(\ln \eta
) $ by means of the decomposition (\ref{decompo}) and from that we obtain $%
\tilde{Q}_{\text{t}}(\alpha _{k})=Q_{\text{t}}(0)$. The outcome is 
\begin{eqnarray}
\tilde{Q}_{\text{t}}(\alpha _{k})=\frac{i}{\sqrt{2}} &&\left[ 1+\frac{3\zeta
^{2}\alpha _{k}^{2}}{4}(8+7\xi )-3\zeta \alpha _{k}^{2}\tilde{\gamma}%
_{M}-\zeta \alpha _{k}^{3}(6\tilde{\gamma}_{M}^{2}+\pi ^{2})\right.  \notag
\\
&&\left. +\frac{3\zeta ^{2}\alpha _{k}^{3}}{2}\tilde{\gamma}_{M}(8+\xi )+%
\frac{3\xi \zeta ^{3}\alpha _{k}^{3}}{8}(22+41\xi +6\xi ^{2})+\mathcal{O}%
(\alpha _{k}^{4})\right] .  \label{Q}
\end{eqnarray}

Inserting this result into (\ref{ptQ}) we finally obtain the spectrum to the
NNLL order, which reads 
\begin{eqnarray}
\mathcal{P}_{T}\left( k\right) =\frac{4m_{\phi }^{2}\zeta G}{\pi } &&\left[
1-3\zeta \alpha _{k}\left( 1+2\alpha _{k}\gamma _{M}+4\gamma _{M}^{2}\alpha
_{k}^{2}-\frac{\pi ^{2}\alpha _{k}^{2}}{3}\right) +\frac{\zeta ^{2}\alpha
_{k}^{2}}{8}(94+11\xi )\right.  \notag \\
&&\left. +3\gamma _{M}\zeta ^{2}\alpha _{k}^{3}(14+\xi )-\frac{\zeta
^{3}\alpha _{k}^{3}}{12}(614+191\xi +23\xi ^{2})+\mathcal{O}(\alpha _{k}^{4})%
\right] .\qquad  \label{ptF}
\end{eqnarray}

\subsection{Checks of the RG\ equation and the spectrum}

\label{checks}

Now we perform two checks of the RG equation and one of the spectrum. The
first check of the RG equation is straightforward. Indeed, the expression of 
$Q_{\text{t}}(\ln \eta )$ just found by solving (\ref{weq}) must coincide
with the RG improved one -- given by formula (\ref{Qtt}) with the overall
constant (\ref{Q}) --, once we ignore the RG improvement (that is to say, to
order $\alpha _{k}^{3}$ and considering $\alpha _{k}\ln \eta $ of order one
instead of order zero). Using the expansions (\ref{beha}) of the appendix
and the relations (\ref{wm}), we easily verify that it is so.

We wish to make a more invasive, independent check of the RG equation to the
NNLL order. To achieve this goal, which also provides a check of the
spectrum, we need the projected $U$ Lagrangian to order $\alpha ^{4}$, which
we do not have exactly. Since the projection $V(U)$ is obtained by
integrating out $V$ with the fakeon prescription, the solution $V(U)=%
\mathcal{O}(\alpha ^{2})$ is nonlocal, in general. This feature of $V(U)$
becomes visible beyond the superhorizon limit and before using the $U$ field
equations. The scale of nonlocality is the fakeon mass $m_{\chi }$ and is
also the scale of the violation of microcausality \cite{classicization,FLRW}%
. The second and third lines of (\ref{ltg}) show that the projected $U$
Lagrangian is nonlocal starting from order $\alpha ^{4}$. Yet, we know that
these issues do not affect the RG improved spectrum to the NNLL\ order. One
way to bypass the difficulty is to take $m_{\chi }$ large by expanding in
powers of $\xi $ and realizing that the lowest orders of the expansion give
the exact result. In the derivation below, we manage to check the RG
equation and the spectrum to order $\xi ^{5}$.

Moving the subleading terms to the right-hand side, the $V$ field equation
can be arranged into the form%
\begin{equation}
V=\xi \alpha ^{2}\left( v_{4}+v_{5}\alpha +\frac{(v_{6}+v_{7}\alpha )\xi
k^{2}}{m_{\phi }^{2}a^{2}}\right) U-\frac{\xi }{2}\left( 1-3\alpha +\frac{%
2k^{2}}{m_{\phi }^{2}a^{2}}\right) V-\frac{\xi \ddot{V}}{m_{\phi }^{2}}-%
\frac{3\xi H\dot{V}}{m_{\phi }^{2}}  \label{Veq}
\end{equation}%
and solved perturbatively in $\xi $, where $v_{i}$,\ $i=4,\ldots 7,$ are
constants. The solution $V(U)$ is then inserted back into $\mathcal{L}_{%
\text{t}}^{\prime }(U,V)$, to obtain the projected Lagrangian $\mathcal{L}_{%
\text{t}}^{\prime \prime }(U)=\mathcal{L}_{\text{t}}^{\prime }(U,V(U))$. Due
to the expansion in powers of $\xi $, $\mathcal{L}_{\text{t}}^{\prime \prime
}$ is local, but contains higher derivatives of $U$ and high powers of $%
k^{2}/(m_{\phi }^{2}a^{2})$. We eliminate both by means of a field
redefinition%
\begin{equation}
U(W)=W+\xi ^{2}\alpha ^{4}\sum_{j\geqslant 0}\tilde{h}_{j}(\xi )W^{(j)},
\label{UW}
\end{equation}%
such that $\mathcal{L}_{\text{t}}^{\prime \prime \prime }(W)=\mathcal{L}_{%
\text{t}}^{\prime \prime }(U(W))$ is cast into the standard form%
\begin{equation}
(8\pi G)\frac{\mathcal{L}_{\text{t}}^{\prime \prime \prime }(W)}{a^{3}}=\dot{%
W}^{2}-\frac{k^{2}}{a^{2}}W^{2}+\xi \alpha ^{2}\left[ h_{1}(\alpha ,\xi )+%
\frac{k^{2}}{a^{2}}h_{2}(\alpha ,\xi )\right] W^{2},  \label{LE}
\end{equation}%
where $h_{1,2}(\alpha ,\xi )$ are power series in $\alpha $ and $\xi $ and $%
\tilde{h}_{j}(\xi )$ are power series in $\xi $.

Finally, defining, $w=a\sqrt{k}W/\sqrt{4\pi G}$, formula (\ref{LE}) gives
the projected $w$ action, which turns out to have the form (\ref{sredF})
with the same $h$ (to order $\alpha ^{3}$), but a different $\sigma _{\text{t%
}}$. Note that various field redefinitions are involved, so only the final
results (the spectra $\mathcal{P}$) should match. The ingredients of the
intermediate steps do not need to coincide. This means that we must repeat
the derivation to the very end with the new $\sigma _{\text{t}}$: first, we
work out (\ref{Qtt}) from (\ref{trun}); second, we extract $\tilde{Q}_{\text{%
t}}(\alpha _{k})=Q_{\text{t}}(0)$ from the solution to the $w$ equation;
third, we set $\eta w=$ $Q_{\text{t}}(\ln \eta )=\tilde{Q}_{\text{t}}(\alpha
,\alpha _{k})$ in the superhorizon limit, according to (\ref{decompo}); then
we have $W=\sqrt{4\pi G}w/(a\sqrt{k})$, which allows us to obtain $U$ from (%
\ref{UW}), $V$ from the recursive solution to (\ref{Veq}), $u$ from (\ref{uU}%
) and finally the spectrum from (\ref{pt}). At the end, we correctly find
the expansion of (\ref{ptF}) in powers of $\xi $, which we push to order $%
\xi ^{5}$.

The steps described above allow us to calculate the RG improved spectrum
directly, while the arguments of the previous subsections give $F_{\text{t}}$%
, and hence $\mathcal{P}_{T}$, \textit{from} the RG equation. The
approximation to order $\xi ^{5}$ is also able to determine $F_{\text{t}}$
exactly, because it outperforms $F_{\text{t}}$, which is a polynomial of
degree three in $\xi $. The results we have just obtained provide the
desired checks of the RG equation and the spectrum.

\section{Quantum gravity:\ scalar fluctuations}

\label{scalarocm}

In this section we derive the running scalar spectrum to the NNLL order,
which is the first order affected by the fakeon $\chi _{\mu \nu }$ in
quantum gravity. We do not calculate the NNLL corrections exactly in $\xi $,
because they lead to a nonlocal projected action. Instead, we expand in
powers of $\xi $. The result contains an asymptotic series that we work out
to order $\xi ^{9}$.

We expand the Lagrangian (\ref{sqgeq}) by means of (\ref{mets}) (with $u=v=0$%
)\ to the quadratic order in the fluctuations. Then we eliminate the
auxiliary field $\Phi $ by means of its own field equation. Finally, we make
the field redefinitions%
\begin{equation}
\Psi =\frac{U}{\sqrt{3}\alpha \sqrt{1+\alpha ^{2}}},\qquad B=\frac{a^{2}}{%
k^{2}}V+\frac{U}{\sqrt{3}\alpha H(1-\alpha ^{2})\sqrt{1+\alpha ^{2}}}.
\label{psib}
\end{equation}%
So doing, we obtain a Lagrangian $\mathcal{L}_{\text{s}}^{\prime }(U,V)$
that admits a regular expansion in powers of $k$ and $\alpha $. We do not
report its expression here, because it is rather involved.

The next step is to determine the fakeon projection $V(U)$, which starts
from order $\alpha $. For the purposes of this paper, it is sufficient to
work out the projected Lagrangian $\mathcal{L}_{\text{s}}^{\prime \prime
}(U)=\mathcal{L}_{\text{s}}^{\prime }(U,V(U))$ to order $\alpha ^{3}$
included, which requires to study $V(U)$ to order $\alpha ^{2}$.

As said, we perform the projection by expanding in powers of $\xi =m_{\phi
}^{2}/m_{\chi }^{2}$ and solving the $V$ equation of motion iteratively in $%
\xi $. We first illustrate the procedure by writing the key formulas to the
first order. Later we describe how to handle the higher orders.

To order $\xi $, we find%
\begin{eqnarray}
V(U) &=&-\frac{\sqrt{3}}{2}m_{\phi }\left[ (2-\xi )\alpha +\frac{2}{3}%
(2+3\alpha +4\xi \alpha )\frac{k^{2}}{m_{\phi }^{2}a^{2}}-\frac{2\xi
(2+3\alpha )}{3}\frac{k^{4}}{m_{\phi }^{4}a^{4}}\right] \alpha U  \notag \\
&&+\frac{\sqrt{3}}{2}\left( 2-\xi +3\xi \alpha -\frac{4\xi }{3}\frac{k^{2}}{%
m_{\phi }^{2}a^{2}}\right) \alpha \dot{U}  \label{Vv} \\
&&-\frac{\sqrt{3}}{2}\left( 3-\frac{5}{2}\alpha -\frac{2}{3}(2+3\alpha )%
\frac{k^{2}}{m_{\phi }^{2}a^{2}}\right) \frac{\xi \alpha }{m_{\phi }}\ddot{U}%
-\frac{\sqrt{3}\xi \alpha U^{(3)}}{m_{\phi }^{2}}+\mathcal{O}(\alpha ^{3}). 
\notag
\end{eqnarray}%
Inserting the solution back into $\mathcal{L}_{\text{s}}^{\prime }(U,V)$, we
obtain the projected Lagrangian $\mathcal{L}_{\text{s}}^{\prime \prime }(U)$%
, which reads%
\begin{eqnarray}
(8\pi G)\frac{\mathcal{L}_{\text{s}}^{\prime \prime }}{a^{3}} &=&\left[
1-\alpha ^{2}+\frac{\xi }{2}\alpha ^{2}(1-3\alpha )+\frac{2\xi \alpha
^{2}k^{2}}{m_{\phi }^{2}a^{2}}\right] \dot{U}^{2}-\frac{k^{2}}{a^{2}}\left(
1-\alpha ^{2}+\frac{\xi \alpha ^{2}k^{2}}{m_{\phi }^{2}a^{2}}\right) U^{2} 
\notag \\
&&+\frac{3m_{\phi }^{2}\alpha }{2}\left( 1-\frac{5}{6}\alpha +\frac{29+18\xi 
}{36}\alpha ^{2}\right) U^{2}-\frac{\xi \alpha ^{2}\ddot{U}^{2}}{m_{\phi
}^{2}}+\mathcal{O}(\alpha ^{4}).  \label{hdscal}
\end{eqnarray}%
The higher-derivatives can be eliminated by means of the change of variables%
\begin{equation}
U=u\left( 1+\frac{\alpha ^{2}}{2}\right) -\frac{\xi \alpha ^{2}}{4}\left(
1+6\alpha +\frac{2k^{2}}{m_{\phi }^{2}a^{2}}\right) u-\frac{\xi \alpha
^{2}(6+23\alpha )\dot{u}}{8m_{\phi }}-\frac{\xi \alpha ^{2}\ddot{u}}{%
2m_{\phi }^{2}}+\mathcal{O}(\alpha ^{4}),  \label{Uu}
\end{equation}%
which turns (\ref{hdscal}) into the standard form%
\begin{equation}
(8\pi G\mathcal{)}\frac{\mathcal{L}_{\text{s}}^{\prime \prime \prime }(u)}{%
a^{3}}=\dot{u}^{2}-\frac{k^{2}}{a^{2}}u^{2}+\frac{3m_{\phi }^{2}\alpha }{2}%
\left( 1-\frac{5}{6}\alpha +\frac{29}{36}\alpha ^{2}+\frac{\xi \alpha ^{2}}{2%
}\right) u^{2}.  \label{lau}
\end{equation}

If we want to include higher orders, the procedure does not change by much.
Once $V(U)$ and $\mathcal{L}_{\text{s}}^{\prime \prime }(U)$ are obtained,
the key point is to find the change of variables 
\begin{equation}
U=\left[ 1+\frac{\alpha ^{2}}{2}-\frac{\alpha ^{2}\xi }{4}F_{\text{s}}(\xi )%
\right] u+\alpha ^{2}\mathcal{O}(k^{2}|\tau |^{2},\dot{u},\ddot{u},\cdots )+%
\mathcal{O}(\alpha ^{3})  \label{Uu8}
\end{equation}%
that turns the projected action $\mathcal{L}_{\text{s}}^{\prime \prime
\prime }(u)$ into the form 
\begin{equation}
(8\pi G\mathcal{)}\frac{\mathcal{L}_{\text{s}}^{\prime \prime \prime }(u)}{%
a^{3}}=\dot{u}^{2}-\frac{k^{2}}{a^{2}}u^{2}+m^{2}(\alpha )u^{2},
\label{ls3m}
\end{equation}%
for a suitable $m^{2}(\alpha )=\mathcal{O}(\alpha )$.

For the power spectrum, it is sufficient to report the function $F_{\text{s}%
}(\xi )$ of (\ref{Uu8}). To the eighth order included, we find%
\begin{equation}
F_{\text{s}}(\xi )=1+\frac{\xi }{4}+\frac{\xi ^{2}}{8}+\frac{\xi ^{3}}{8}+%
\frac{7\xi ^{4}}{32}+\frac{19}{32}\xi ^{5}+\frac{295}{128}\xi ^{6}+\frac{1549%
}{128}\xi ^{7}+\frac{42271}{512}\xi ^{8}+\mathcal{O}(\xi ^{9}),  \label{px}
\end{equation}%
which suggests that the series expansion of $F_{\text{s}}(\xi )$ is
asymptotic. It is possible to extend the calculation to arbitrarily high
orders of $\xi $.

Finally, the Lagrangian (\ref{ls3m}) becomes%
\begin{equation}
(8\pi G\mathcal{)}\frac{\mathcal{L}_{\text{s}}^{\prime \prime \prime }(u)}{%
a^{3}}=\dot{u}^{2}-\frac{k^{2}}{a^{2}}u^{2}+\frac{3m_{\phi }^{2}\alpha }{2}%
\left( 1-\frac{5}{6}\alpha +\frac{29}{36}\alpha ^{2}+\frac{\alpha ^{2}\xi }{2%
}F_{\text{s}}(\xi )\right) u^{2}+\mathcal{O}(\alpha ^{4}).  \label{ls3}
\end{equation}

The terms $\mathcal{O}(k^{2},\dot{u},\ddot{u},\cdots )$ and $\mathcal{O}%
(\alpha ^{3})$ of (\ref{Uu8}) are necessary to derive $F_{\text{s}}(\xi )$
and this expression. However, they are rather involved and can be ignored
from this point onwards, so we do not report them here. In particular, we
can drop them when we derive the spectrum to order $\alpha ^{2}$ in the
superhorizon limit $k/(m_{\phi }a)\ll 1$, i.e., $\eta \ll 1$. In this
respect, note that the Lagrangian (\ref{ls3m}) ensures that at $k=0$, $%
\alpha =0$ the solution is $u=$ constant, which implies $\dot{u}=\mathcal{O}%
(\alpha )$ for $k/(m_{\phi }a)\rightarrow 0$.

As far as the Mukhanov-Sasaki action is concerned, we find (\ref{sred}) with%
\begin{equation}
w=\frac{a\sqrt{k}}{\sqrt{4\pi G}}u,\qquad \sigma _{\text{s}}=6\alpha
+22\alpha ^{2}+\left( \frac{280}{3}+3\xi F_{\text{s}}(\xi )\right) \alpha
^{3}+\mathcal{O}(\alpha ^{4}).  \label{ws}
\end{equation}%
Solving (\ref{trun}), we obtain%
\begin{equation}
Q^{(\mathcal{R)}}(\ln \eta )=\tilde{Q}_{\text{s}}(\alpha _{k})\frac{\alpha }{%
\alpha _{k}}\frac{J_{\text{s}}(\alpha )}{J_{\text{s}}(\alpha _{k})},\qquad
J_{\text{s}}(\alpha )=1+\frac{3\alpha }{2}+\left( \frac{7}{2}+\frac{\xi }{4}%
F_{\text{s}}(\xi )\right) \alpha ^{2}+\mathcal{O}(\alpha ^{3}).  \notag
\end{equation}

We go back to $\Psi $ as follows. Formula (\ref{decompo}) tells us that $w$
is equal to $Q/\eta $ plus corrections that are negligible in the
superhorizon limit. Then (\ref{ws}) gives $u$, (\ref{Uu8}) gives $U$ and (%
\ref{psib}) leads to $\Psi $. At last, we use (\ref{pR}) and obtain the
spectrum%
\begin{equation}
\mathcal{P}_{\mathcal{R}}(k)=\frac{Gm_{\phi }^{2}}{6\pi \alpha ^{2}}|\tilde{Q%
}_{\text{s}}(\alpha _{k})|^{2}\left[ 1-3\alpha _{k}-\frac{\alpha _{k}^{2}}{4}%
\left( 1+2\xi F_{\text{s}}(\xi )\right) +\mathcal{O}(\alpha _{k}^{3})\right]
.  \label{PRQ}
\end{equation}%
As expected, the $\alpha $ dependence disappears, in agreement with the RG
equation (\ref{noalfa}).

The last step is to compute $\tilde{Q}_{\text{s}}(\alpha _{k})$, which is
straightforward, since it coincides with the result of (\ref{Q0staro}). The
reason is that the action (\ref{ls3}) depends on $m_{\chi }$ only from order 
$\alpha ^{3}$, so $\tilde{Q}_{\text{s}}(\alpha _{k})$ is unaffected by $%
m_{\chi }$ to order $\alpha _{k}^{2}$. At the end, we get 
\begin{equation}
\mathcal{P}_{\mathcal{R}}(k)=\frac{Gm_{\phi }^{2}}{12\pi \alpha _{k}^{2}}%
\left[ 1+(5-4\gamma _{M})\alpha _{k}+\left( 4\gamma _{M}^{2}-\frac{40}{3}%
\gamma _{M}+\frac{7}{3}\pi ^{2}-\frac{67}{12}-\frac{\xi }{2}F_{\text{s}}(\xi
)\right) \alpha _{k}^{2}+\mathcal{O}(\alpha _{k}^{3})\right] \!\!.
\label{PR}
\end{equation}

\section{Predictions}

\label{predictions}\setcounter{equation}{0}

If primordial cosmology turns into an arena for precision tests of quantum
gravity, as we hope and deem realistic, the predictions of this paper have a
chance to be tested in the incoming years \cite{CMBStage4}. In this section
we summarize them and comment on their validity. Besides the power spectra (%
\ref{ptF}) and (\ref{PR}), a number of other quantities can be calculated
straightforwardly from them. We mention the running (\textquotedblleft
dynamical\textquotedblright ) tensor-to-scalar ratio 
\begin{equation}
r(k)=\frac{\mathcal{P}_{T}(k)}{\mathcal{P}_{\mathcal{R}}(k)}  \label{dynr}
\end{equation}%
as a function of $k$, as well as the tilts%
\begin{equation*}
n_{T}=-\beta _{\alpha }(\alpha _{k})\frac{\partial \ln \mathcal{P}_{T}}{%
\partial \alpha _{k}},\qquad n_{\mathcal{R}}-1=-\beta _{\alpha }(\alpha _{k})%
\frac{\partial \ln \mathcal{P}_{\mathcal{R}}}{\partial \alpha _{k}},
\end{equation*}%
and the running coefficients%
\begin{equation*}
\frac{\mathrm{d}^{n}n_{T}}{\mathrm{d}\ln k\hspace{0.01in}^{n}}=\left( -\beta
_{\alpha }(\alpha _{k})\frac{\partial }{\partial \alpha _{k}}\right)
^{n}n_{T},\qquad \frac{\mathrm{d}^{n}n_{\mathcal{R}}}{\mathrm{d}\ln k\hspace{%
0.01in}^{n}}=\left( -\beta _{\alpha }(\alpha _{k})\frac{\partial }{\partial
\alpha _{k}}\right) ^{n}n_{\mathcal{R}}.
\end{equation*}

Using (\ref{ptF}) and (\ref{PR}) we find%
\begin{eqnarray}
n_{T} &=&-6\left[ 1+4\gamma _{M}\alpha _{k}+(12\gamma _{M}^{2}-\pi
^{2})\alpha _{k}^{2}\right] \zeta \alpha _{k}^{2}+\left[ 24+3\xi +4(31+2\xi
)\gamma _{M}\alpha _{k}\right] \zeta ^{2}\alpha _{k}^{3}  \notag \\
&&\qquad \qquad -\frac{1}{8}(1136+566\xi +107\xi ^{2})\zeta ^{3}\alpha
_{k}^{4}+\mathcal{O}(\alpha _{k}^{5}),  \label{tilts} \\
n_{\mathcal{R}}-1 &=&-4\alpha _{k}+\frac{4\alpha _{k}^{2}}{3}(5-6\gamma
_{M})-\frac{2\alpha _{k}^{3}}{9}(338-90\gamma _{M}+72\gamma _{M}^{2}-42\pi
^{2}+9\xi F_{\text{s}})+\mathcal{O}(\alpha _{k}^{4}).\qquad  \notag
\end{eqnarray}%
The running coefficients follow immediately form the RG equation and the
beta function. The first two corrections to the relation $r+8n_{T}=0$ can be
derived from (\ref{ptF}), (\ref{PR}) and (\ref{tilts}):%
\begin{equation}
r+8n_{T}=-192\zeta \alpha _{k}^{3}+8(202\zeta +65\xi \zeta -144\gamma
_{M}-8\pi ^{2}+3\xi F_{\text{s}})\zeta \alpha _{k}^{4}+\mathcal{O}(\alpha
_{k}^{5}).  \label{ratiodevia}
\end{equation}%
We do not give the expression of $r$ separately, since it can be obtained
straightforwardly by combining (\ref{tilts}) and (\ref{ratiodevia}).

To discuss the validity of the predictions, we express the results in terms
of a pivot\ scale $k_{\ast }$ and evolve $\alpha (1/k)$ from $k_{\ast }$ to $%
k$ by means of the RG evolution equations, using the NNLL running coupling (%
\ref{runca}). So doing, the spectra become functions of $\ln (k_{\ast }/k)$
and the pivot coupling $\alpha _{\ast }\equiv \alpha (1/k_{\ast })$. With $%
k_{\ast }=0.05$ Mpc$^{-1}$ and (for definiteness) $\xi \sim F_{\text{s}}\sim
1$, the data\ reported in \cite{Planck18} give $\ln (10^{10}\mathcal{P}_{%
\mathcal{R}}^{\ast })=3.044\pm 0.014$ and $n_{\mathcal{R}}^{\ast }=0.9649\pm
0.0042$, where the star superscript means that the quantity is evaluated at
the pivot scale. Formulas (\ref{PR}) and (\ref{tilts}) then give the values $%
\alpha _{\ast }=0.0087\pm 0.0010$ and $m_{\phi }=(2.99\pm 0.37)\cdot 10^{13}$%
GeV for the fine structure constant $\alpha _{\ast }$ and the inflaton mass,
respectively.

The value of $m_{\chi }$ will be known as soon as the tensor-to-scalar ratio 
$r$ will be measured. We recall that the mass $m_{\chi }$ of the fakeon $%
\chi _{\mu \nu }$ is constrained to lie in the range $m_{\phi }/4<m_{\chi
}\,<\infty $, which means $0<\xi <16$, by the consistency of the fakeon
prescription/projection \cite{ABP}. This restricts the window of allowed
values of $r$ to $4\cdot 10^{-4}\lesssim r\lesssim 3.5\cdot 10^{-3}$ at the
pivot scale.

Formula (\ref{ptF}) predicts the tensor spectrum $\mathcal{P}_{T}$ with a
relative theoretical error equal to $\alpha _{\ast }^{4}\sim 10^{-8}$, while
the relative error on the tensor tilt $n_{T}$ is $\alpha _{\ast }^{3}\sim
10^{-6}$. As far as the quantities involving the scalar fluctuations are
concerned, we have to take into account that the function $F_{\text{s}}(\xi
) $ is only partially known. To find when the expansion (\ref{px}) of $F_{%
\text{s}}(\xi )$ can be trusted, and estimate its errors, we proceed as
follows.

Typically, an asymptotic series is a sum of corrections that decrease up to
a certain point and then blow up. The expansion must be truncated right at
that point, because it cannot be trusted further. The last good term,
multiplied by $\pm 1/2$, can be used to give an estimate of the error. From (%
\ref{px}) we see that for $4<\xi <16$, we can trust no correction at all, so
we can say $F_{\text{s}}(\xi )=1\pm 1$ for $\xi >4$. For $2<\xi <4$ the
second term is smaller than the first one, but also smaller than the third
one, so $F_{\text{s}}(\xi )=1+\xi /4\pm \xi /8$. For $1<\xi <2$ the third
term is smaller than the second one as well as the fourth one, so $F_{\text{s%
}}(\xi )=1+\xi /4+\xi ^{2}/8\pm \xi ^{2}/16$. And so on: 
\begin{eqnarray*}
F_{\text{s}}(\xi ) &=&1+\frac{\xi }{4}+\frac{\xi ^{2}}{8}+\frac{\xi ^{3}}{8}%
\pm \frac{\xi ^{3}}{16}\qquad \text{for }\frac{4}{7}<\xi <1, \\
F_{\text{s}}(\xi ) &=&1+\frac{\xi }{4}+\frac{\xi ^{2}}{8}+\frac{\xi ^{3}}{8}+%
\frac{7\xi ^{4}}{32}\pm \frac{7\xi ^{4}}{64}\qquad \text{for }\frac{7}{19}%
<\xi <\frac{4}{7},
\end{eqnarray*}%
etc. For example, we have $F_{\text{s}}(1/2)=1.186\pm 0.007$. An error
around 1\% is comparable with the error we are making anyway by neglecting
the NNNLL order. This means that the NNLL prediction on the scalar spectrum,
obtained by expanding in powers of $\xi $, is good for every $\xi <1/2$.
Even for $1/2<\xi <1$ (where the error is below 6\%) it is a fair
improvement with respect to the NLL prediction.

Summarizing, the scalar spectrum $\mathcal{P}_{\mathcal{R}}$ and the scalar
tilt $n_{\mathcal{R}}-1$ have a relative theoretical error that can be
estimated around $\alpha _{\ast }^{3}\sim 10^{-6}$ for $\xi <1/2$, where the
expansion in powers of $\xi $ is most successful. In the remaining interval
of values of $\xi $, the relative error ranges from $10^{-5}$ (for $1/2<\xi
<1$) to $10^{-4}$ (for $1<\xi <16$).

\section{Conclusions}

\label{conclusions}\setcounter{equation}{0}

Primordial cosmology gives us the chance to test quantum gravity in the
forthcoming future. The results may trigger a virtuous circle and open a
season of precision tests, which can hopefully lead to the same level of
success experienced by the standard model of particle physics. To make this
possible, it is crucial to overcome the lack of predictivity due to the
arbitrariness of classical theories. The constraints of quantum field theory
allow us to achieve this goal, by singling out a theory that contains very
few parameters. Additional constraints coming from cosmology make its
predictions quite sharp, starting from the tensor-to-scalar ratio $r$.

Various similarities with high-energy physics allow us to import techniques
from quantum field theory into cosmology and boost the computations of
higher-order corrections. The cosmic RG flow, for example, allows us to
enhance the spectra of primordial fluctuations and understand their
structure better. In this paper, we have computed the RG improved spectra of
the curvature perturbation $\mathcal{R}$ and the tensor fluctuations,
together with the quantity $r+8n_{T}$, to the NNLL\ order in the
superhorizon limit. Tilts and running coefficients follow straightforwardly
from the flow. The relative theoretical errors range from $\alpha ^{4}\sim
10^{-8}$ to $\alpha ^{3}\sim 10^{-6}$.

We have made independent checks of the cosmic RG equations. The results show
that the tensor spectrum is affected by the fakeon $\chi _{\mu \nu }$ at
every order of the expansion in $\alpha $. As far as the scalar spectrum is
concerned, the first corrections that depend on $m_{\chi }$ are the NNLL
ones, which we have computed by expanding perturbatively in $m_{\phi
}/m_{\chi }$, due to nontrivial issues related with the fakeon projection at
high orders. This approach gives precise results for $m_{\chi }^{2}>2m_{\phi
}^{2}$ and fairly precise ones for $2m_{\phi }^{2}>m_{\chi }^{2}>m_{\phi
}^{2}$.

So far, we have been able to treat the fakeon projection on nontrivial
backgrounds relatively easily. It was far from obvious that we could do so
to such high orders. From the practical point of view, it may be unnecessary
to push the calculations further, at least for the moment. Yet, it is
conceptually interesting to understand how to handle the corrections
systematically to arbitrary orders. Without entering into details, a
preliminary analysis allows us to anticipate that the fakeon projection 
\textit{per se}, which can be defined algorithmically, does not pose the
major challenge. What appears to be more difficult is to properly generalize
the Bunch-Davies vacuum condition. A deeper investigation into these issues
may clarify the missing points.

\vskip10truept \noindent {\large \textbf{Acknowledgments}}

\vskip 1truept

We thank M. Piva for helpful discussions and NICPB, Tallin, Estonia, for
hospitality during the final stage of this work.

\renewcommand{\thesection}{\Alph{section}} \renewcommand{\theequation}{%
\thesection.\arabic{equation}} \setcounter{section}{0}

\section{Appendix. Reference formulas}

\label{formulas} \setcounter{equation}{0}

In this appendix we collect some formulas used in the paper. We start from
the running coupling to the NNLL order, from \cite{CMBrunning}, which reads%
\begin{equation}
\alpha =\frac{\alpha _{k}}{\lambda }\left( 1-\frac{5\alpha _{k}}{6\lambda }%
\ln \lambda \right) \left[ 1+\frac{25\alpha _{k}^{2}}{12\lambda ^{2}}\left(
1-\lambda -\frac{\ln \lambda }{3}(1-\ln \lambda )\right) \right] ,
\label{runca}
\end{equation}%
where $\lambda \equiv 1+2\alpha _{k}\ln \eta $, $\eta =-k\tau $. We also
have the expansions%
\begin{eqnarray}
H &=&\frac{m_{\phi }}{2}\left[ 1-\frac{3\alpha }{2}+\frac{7\alpha ^{2}}{4}-%
\frac{47\alpha ^{3}}{24}+\frac{293\alpha ^{4}}{144}-\frac{1645}{864}\alpha
^{5}+\frac{5489}{5184}\alpha ^{6}+\mathcal{O}(\alpha ^{7})\right] ,
\label{acca} \\
-aH\tau &=&1+3\alpha ^{2}+12\alpha ^{3}+91\alpha ^{4}+\frac{2464}{3}\alpha
^{5}+\frac{81659}{9}\alpha ^{6}+\mathcal{O}(\alpha ^{7}).  \label{aHtau}
\end{eqnarray}

Next, we give the functions $g_{n}(\eta )$ of equations (\ref{weqsstaro})
and their solutions $w_{n}(\eta )$, determined by the Bunch-Davies condition
(\ref{bunch}). For the tensor perturbations in the limit of infinitely heavy
fakeon, we have $g_{0}^{\text{t}}=g_{1}^{\text{t}}=0$, $g_{2}^{\text{t}%
}=9w_{0}^{\text{t}}$, $g_{3}^{\text{t}}=12w_{0}^{\text{t}}(4-3\ln \eta )$,
with solutions 
\begin{eqnarray}
&&w_{0}^{\text{t}}=\frac{i(1-i\eta )}{\eta \sqrt{2}}\mathrm{e}^{i\eta
},\qquad w_{1}^{\text{t}}=0,\qquad w_{2}^{\text{t}}=\frac{6w_{0}^{\text{t}}}{%
1-i\eta }-3\left( i\pi -\hspace{0.01in}\text{Ei}(2i\eta )\right) w_{0}^{%
\text{t}\ast },  \notag \\
&&w_{3}^{\text{t}}=\left[ 6(\ln \eta +\tilde{\gamma}_{M})^{2}+24i\eta
F_{2,2,2}^{1,1,1}\left( 2i\eta \right) +\pi ^{2}\right] w_{0}^{\text{t}\ast
}+\frac{24w_{0}^{\text{t}}}{1-i\eta }-4(\ln \eta +1)w_{2}^{\text{t}},
\label{wt}
\end{eqnarray}%
where Ei denotes the exponential-integral function and $F_{b_{1},\cdots
,b_{q}}^{a_{1},\cdots ,a_{p}}(z)$ denotes the generalized hypergeometric
function $_{p}F_{q}(\{a_{1},\cdots ,a_{p}\},\{b_{1},\cdots ,b_{q}\};z)$.

We can use the functions (\ref{wt}) to express the solutions in other
situations. Using self-explanatory superscripts to distinguish different
cases, the scalar fluctuations at $m_{\chi }=\infty $ have $g_{0}^{\text{s}%
}=0$, $g_{1}^{\text{s}}=6w_{0}^{\text{s}}$, $g_{2}^{\text{s}}=2(11-6\ln \eta
)w_{0}^{\text{s}}+6w_{1}^{\text{s}}$, with solutions $w_{0}^{\text{s}%
}=w_{0}^{\text{t}}$ and%
\begin{equation*}
w_{1}^{\text{s}}=\frac{2}{3}w_{2}^{\text{t}},\qquad w_{2}^{\text{s}}=-\frac{%
16w_{0}^{\text{t}}}{1+\eta ^{2}}+\frac{2(13+i\eta )w_{2}^{\text{t}}}{%
9(1+i\eta )}+\frac{w_{3}^{\text{t}}}{3}+4G_{2,3}^{3,1}\left( -2i\eta
\left\vert _{0,0,0}^{\ 0,1}\right. \right) w_{0}^{\text{t}},
\end{equation*}%
where $G_{p,q}^{m,n}$ denotes the Meijer G function.

When $m_{\chi }$ is finite, the solutions do not change in the case of the
scalar perturbations. They do change in the case of the tensor
perturbations, where the equations acquire the form (\ref{weq}), instead of (%
\ref{weqsstaro}). It is convenient to organize them as 
\begin{equation}
w_{n}^{\prime \prime }+w_{n}-\frac{2w_{n}}{\eta ^{2}}=\frac{g_{n}(\eta )}{%
\eta ^{2}}+\tilde{g}_{n}(\eta ),  \label{weqm}
\end{equation}%
where the functions $g_{n}$ and $\tilde{g}_{n}$ are determined recursively
from $w_{m}$, $m<n$. We find $g_{0}^{\text{t},m_{\chi }}=g_{1}^{\text{t}%
,m_{\chi }}=\tilde{g}_{0}^{\text{t},m_{\chi }}=\tilde{g}_{1}^{\text{t}%
,m_{\chi }}=0$ and%
\begin{eqnarray}
g_{2}^{\text{t},m_{\chi }} &=&9\zeta w_{0}^{\text{t},m_{\chi }},\qquad 
\tilde{g}_{2}^{\text{t},m_{\chi }}=3\xi \zeta ^{2}w_{0}^{\text{t},m_{\chi
}},\qquad g_{3}^{\text{t},m_{\chi }}=\frac{3}{2}(32\zeta +43\xi \zeta -24\ln
\eta )\zeta w_{0}^{\text{t},m_{\chi }},  \notag \\
\tilde{g}_{3}^{\text{t},m_{\chi }} &=&-\frac{3}{2}(6\zeta -7\xi \zeta -2\xi
^{2}\zeta +8\ln \eta )\xi \zeta ^{2}w_{0}^{\text{t},m_{\chi }},  \label{gm}
\end{eqnarray}%
with solutions $w_{0}^{\text{t},m_{\chi }}=w_{0}^{\text{t}}$ and%
\begin{eqnarray}
w_{2}^{\text{t},m_{\chi }} &=&\zeta w_{2}^{\text{t}}+\frac{3\xi \zeta ^{2}}{4%
}\frac{3-3i\eta -2\eta ^{2}}{1-i\eta }w_{0}^{\text{t}},\qquad w_{3}^{\text{t}%
,m_{\chi }}=\zeta w_{3}^{\text{t}}+\frac{3\xi \zeta ^{2}}{2}w_{2}^{\text{t}}+%
\frac{3\xi \zeta ^{2}F_{3}w_{0}^{\text{t}}}{4(1-i\eta )},  \notag \\
F_{3} &\equiv &(3-3i\eta -2\eta ^{2})(\xi ^{2}\zeta -4\ln \eta )-\zeta
(1+i\eta )(1-2i\eta )+\frac{\xi \zeta }{2}(29-29i\eta -22\eta ^{2}).\qquad
\label{wm}
\end{eqnarray}

Now we give details on the asymptotic behaviors. The functions $w_{n}(\eta )$
with $n>0$ tend to zero for $\eta \rightarrow \infty $. The relevant
behaviors in the superhorizon limit $\eta \sim 0$ are%
\begin{eqnarray}
\eta w_{2}^{\text{t}} &\sim &\frac{3i}{\sqrt{2}}\left( 2-\tilde{\gamma}%
_{M}-\ln \eta \right) ,\qquad \eta w_{2}^{\text{s}}\sim -8i\sqrt{2}+\frac{26%
}{9}\eta w_{2}^{\text{t}}+\frac{\eta w_{3}^{\text{t}}}{3}+i\sqrt{2}(\ln \eta
+\tilde{\gamma}_{M})^{2}+\frac{i\pi ^{2}}{\sqrt{2}},  \notag \\
\eta w_{3}^{\text{t}} &\sim &-3i\sqrt{2}(\ln \eta +\tilde{\gamma}_{M})^{2}-%
\frac{i\pi ^{2}}{\sqrt{2}}+12i\sqrt{2}-4(\ln \eta +1)\eta w_{2}^{\text{t}}.
\label{beha}
\end{eqnarray}%
The other behaviors can be found from these ones and the relations given
above.

For convenience, we conclude with a list of notations frequently used in the
paper, which are%
\begin{equation*}
\xi =\frac{m_{\phi }^{2}}{m_{\chi }^{2}},\qquad \zeta =\left( 1+\frac{\xi }{2%
}\right) ^{-1},\qquad \tilde{\gamma}_{M}=\gamma _{M}-\frac{i\pi }{2},\qquad
\gamma _{M}=\gamma _{E}+\ln 2,
\end{equation*}%
$\gamma _{E}$ being the Euler-Mascheroni constant.

\end{document}